\documentclass[aps,prd,twocolumn,superscriptaddress,longbibliography,nofootinbib]{revtex4-2}
\usepackage[utf8]{inputenc}
\usepackage{color}
\usepackage{graphicx}
\usepackage{subfigure}
\usepackage{xcolor}
\usepackage[normalem]{ulem}
\usepackage[pdftex,breaklinks,colorlinks,linkcolor=blue,citecolor=teal,anchorcolor=red,urlcolor=cyan]{hyperref}
\usepackage{orcidlink}

\usepackage{amsmath,amssymb,amsfonts}

\def\prd{Phys. Rev. D}

\def\apj{Astrophys. J.}

\def\apjl{Astrophys. J. Lett.}
\def\aap{Astronomy and Astrophysics}
\def\pr{Phys. Rev.}

\def\pau_p{Prog. Theor. Phys.}

\def\mnras{Mon. Not. R. Astron. Soc.}

\def\nat{Nature}
\def\sovast{Soviet Ast.}
\def\jcap{J. Cosmology Astropart. Phys}

\def\varv{v}

\begin{document}

\title{Could long-period transients be powered by primordial black hole capture?}

\author{Thomas W.~Baumgarte\orcidlink{0000-0002-6316-602X}}
\affiliation{Department of Physics and Astronomy, Bowdoin College, Brunswick, ME 04011, USA}

\author{Stuart L.~Shapiro\orcidlink{0000-0002-3263-7386}}
\affiliation{Department of Physics, University of Illinois at Urbana-Champaign, Urbana, IL 61801}

\affiliation{Department of Astronomy and NCSA, University of Illinois at Urbana-Champaign, Urbana, IL 61801}

\begin{abstract}
Long-period radio transients have unusual properties that challenge their interpretation as pulsars or magnetars.  We examine whether they might instead be powered by primordial black holes (PBHs) making repeated passages through a host star, thereby providing a signature of elusive dark-matter candidates.  We demonstrate that constraints derived from the transients' period and period derivative alone already rule out this scenario for most potential host stars.  While white dwarfs may satisfy these constraints, they are unlikely to capture PBHs in the required mass range.
\end{abstract}

\maketitle

Analysis of archival data recently revealed the existence of {\em long-period radio transients} (LPRTs; see, e.g., \cite{Caletal22,HWetal22,HWetal23}).  Individual pulses vary significantly in brightness, last for ten to thousands of seconds, and some comprise bursts of significantly shorter duration.  The coherent radio pulses have an unusually long period of up to approximately 20 minutes.  Since some of the archival data go back about three decades, it has also been possible to constrain changes in the period to very small values (see Table \ref{tab:objects} below for the two LPRTs with the longest known periods).  While many of the characteristics of these observations point to pulsars or magnetars as the sources of these pulses, their long periods and low upper limits to their period derivatives in particular challenges this interpretation.  While highly magnetic white dwarfs may have long spin periods,  no close-by magnetic white dwarfs have been observed to generate such emission (see \cite{HWetal22,HWetal23} for detailed discussions).   Hence the origin of these transients remains somewhat mysterious, and we are invited to speculate on more exotic mechanisms that might power LPRTs.

{\em Primordial black holes} (PBHs), first proposed by \cite{ZelN67} and \cite{Haw71}, are hypothetical objects that may have formed in the early Universe (see also \cite{CarH74}).  PBHs have received significant attraction in recent years as potential constituents of the Universe's dark-matter content (see, e.g., \cite{Khl10,CarK20} for reviews; see also \cite{MonCFVSH19}).  They also have been invoked as possible sources of other astronomical phenomena, including the 1908 Tunguska event in Siberia \cite{JakR73} (but see \cite{BeaT74}), neutron star implosions and ``quiet supernovae" \cite{BraL18}, fast radio bursts \cite{AbrBW18}, the formation of low-mass stellar black holes \cite{OncMGG22}, and the origin of supermassive black holes (e.g.~\cite{BamSDFV09}), possibly via the formation of PBH clusters \cite{BelDEGGKKRS14,BelDEEKKKNRS19}.  Gravitational-wave signatures of PBHs have been surveyed recently in \cite{Bagetal23}, and the possibility of using solar-system ephemerides to detect PBHs has been discussed in \cite{BerCDVC23,TraGLK23} and references therein.  While PBHs have been ruled out in some mass ranges, others remain viable.  One such mass window extends from about $10^{-16} M_\odot$ to about $10^{-10} M_\odot$  (labelled ``A" in Fig.~1 of \cite{CarK20}; see also Fig.~10 in \cite{CarKSY21}), where the lower limit results from black-hole evaporation via Hawking radiation in the age of the Universe \citep{Haw74}. 

If PBHs contribute to the dark-matter content of the Universe, some of them are likely to collide with stellar objects (see, e.g., \cite{Abretal09,CapPT13,GenST20} for detailed discussions, estimates, and references).  Inside the star, the PBH experiences drag forces that result from dynamical friction and accretion.  Assuming that the PBH moves at supersonic speeds, this drag force is likely to be dominated by the former (see, e.g., Fig.~2 in \cite{GenST20}) and can be estimated from
\begin{equation} \label{drag}
F_{\rm drag} \simeq  \frac{4 \pi G^2 \rho_* m^2}{\varv^2} \ln \Lambda
\end{equation}
(see \cite{Cha43,RudS71,Ost99}), where $m$ is the PBH's mass, $\varv$ its speed, $\rho_*$ the stellar density, and $\ln \Lambda$ is the Coulombic logarithm that many authors have estimated to be of order ten (e.g.~\cite{Abretal09}).  Further approximating $\rho_* \simeq 3 M_* / (4 \pi R_*^3)$, where $M_*$ and $R_*$ are the stellar mass and radius, and assuming that the PBH passes through the bulk of the star (making it more likely for the PBH to become bound), we see that it will lose an energy
\begin{equation} \label{deltaE}
\Delta E_{\rm drag} \simeq 2 R_* F_{\rm drag} 
\simeq \frac{6 G^2 M_* m^2}{\varv^2 R_*^2} \ln \Lambda
\gtrsim \frac{3 G m^2}{R_*} \ln \Lambda 
\end{equation}
during one passage.  In the last estimate in (\ref{deltaE}) we assumed that the PBH's speed is comparable to or less than the escape speed from the stellar surface, $\varv \lesssim \sqrt{2 G M_* / R_*}$, which is a reasonable approximation assuming that the PBH is captured from a wide orbit about the star.    

Additional energy loss due to the emission of gravitational radiation can be estimated from 
\begin{equation}
    \Delta E_{\rm GW} \simeq \frac{8}{15} \frac{G^{7/2}}{c^5} \frac{m^2 M_*^2 (m + M_*)^{1/2}}{r_{\rm min}^{7/2}} g(e), 
\end{equation}
where $r_{\rm min}$ is the distance of closest approach and $g(e)$ is a dimensionless function of the eccentricity $e$ whose exact form depends on whether the orbit is bound or unbound (see \cite{PetM63,Pet64,Tur77}).  Approximating $r_{\rm min} \simeq R_*$ and assuming $m \ll M_*$ we estimate
\begin{equation} \label{energy_ratio}
    \frac{\Delta E_{\rm GW}}{\Delta E_{\rm drag}} \simeq
    \frac{8 g(e)}{45 \ln \Lambda} \left(\frac{GM_*}{c^2 R_*}\right)^{5/2}.
\end{equation}
Even though $g(e)$ can reach values of $\sim 10^2$ \citep{Tur77}, we see that $\Delta E_{\rm GW}$ can be neglected except possibly for neutron stars.

\begin{table*}
    \centering
    \begin{tabular}{c|c|c|c}
        Object & $P$ & $|\dot P|_{\rm max}$ & Reference \\
        \hline
         GLEAM-XJ162759.5-523504.3 & $1.091 \times 10^3$ s & $1.2 \times 10^{-9}$ & \cite{HWetal22} \\
         GPMJ1839-10 & $1.318 \times 10^3$ s & $3.6 \times 10^{-13}$ & \cite{HWetal23} 
    \end{tabular}
    \caption{Period $P$ and limits on changes in the period $|\dot P|_{\rm max}$ for two recently discovered LPRTs.}
    \label{tab:objects}
\end{table*}

If the loss of energy (\ref{deltaE}) is sufficiently large in a first encounter, the PBH ends up being gravitationally bound to the star.  As long as $m \ll M_*$ it will still emerge from the star, but will ultimately return and potentially complete many more passages through the star before ultimately settling down in its interior and accreting its entire mass (see \cite{EasL19,RicBS21b,SchBS21} for numerical simulations).

In this short Letter we explore whether a PBH in this capture phase, while it makes repeated passages through a host star separated by segments of Keplerian orbits in the stellar exterior, could act as the periodic source of an LPRT.  During each passage the PBH deposits an energy (\ref{deltaE}) to the star that may explain the pulses themselves.  More likely, the PBH will trigger a minor tidal and/or magnetic disturbance at the surface
or atmosphere, or may cause the release of additional nuclear or chemical energy there, each time upon entering and/or exiting the star. Any energy release must ultimately be converted at least in part to radio emission to explain the observations.  (Fortunately, the details of this process will prove unimportant in assessing the viability of this scenario as the source of LPRTs, as we will see below.) For sufficiently small $m$, the PBH's loss of orbital energy would result in only very slow changes in its orbital period, as observed for the two LPRTs listed in Table \ref{tab:objects}.  If such a process were, in fact, the source of LPRTs, it might provide a truly exciting observational signature of PBHs as dark-matter constituents.

Before proceeding we point out a qualitative difference between pulsars and the model that we are considering here.  A pulsar is powered by rotational energy, so that the radiative loss of energy results in a spin-down, and hence an {\em increase} in the pulse period, $\dot P > 0$.  If, on the other hand, the source of energy is orbital energy of an intruder, then the loss of energy leads to a tightening of the orbit, and a {\em decrease} in the pulse period, $\dot P < 0$. A firm observation of the sign of $\dot P$ would therefore distinguish the two different models without ambiguity.  While, for GLEAM-XJ162759.5-523504.3, ``$\dot P > 0$ is preferred" \citep{HWetal22}, a negative $\dot P$ remains possible for both LPRTs listed in Table~\ref{tab:objects}, and we therefore interpret both limits on $\dot P$ as limits on its absolute value $|\dot P|$. 

While the PBH is in relatively wide orbit about the host star, the orbital period, which we identify with the pulse period $P$, is well approximated by Kepler's law
\begin{equation} \label{Kepler}
P^2 = \frac{4 \pi^2}{GM_*} a^3 
\end{equation}
(where $a$ is the orbit's semi-major axis, and where we have again assumed $m \ll M_*$), or
\begin{equation} \label{a}
    a = 1.8 \times 10^{10} \,\mbox{cm} \,\left( \frac{M_*}{M_\odot} \right)^{1/3} \left(\frac{P}{1300 \, \mbox{s}} \right)^{2/3}.
\end{equation}
In the last equation we have scaled $M_*$ to a solar mass $M_\odot$ and the period $P$ to that of GPMJ1839-10, which provides the more stringent limits of the two LPRTs listed in Table~\ref{tab:objects}.  From our discussion above we require that $a > R_*$, which results in our first constraint
\begin{equation} \label{P_constraint}
R_* < 1.8 \times 10^{10}\, \mbox{cm} \,\left( \frac{M_*}{M_\odot} \right)^{1/3} \left(\frac{P}{1300 \, \mbox{s}} \right)^{2/3}.
\end{equation}

We next turn to changes $\dot P$ in the period.  Using Kepler's law (\ref{Kepler}) together with
\begin{equation}
    E = - \frac{G M_* m}{2 a}
\end{equation}
for the orbital energy $E$ we obtain
\begin{equation}
    \frac{dP}{dt} = \frac{6 \pi^2}{GM_*} \frac{a^2}{P} \frac{da}{dt}
    = \frac{12 \pi^2 a^4}{G^2 M^2_* m P} \frac{dE}{dt}.
\end{equation}
We estimate the rate of energy loss from Eq.~(\ref{deltaE}),
\begin{equation}    
    \left| \frac{dE}{dt} \right| \simeq \left| \frac{\Delta E}{P} \right| \gtrsim  \frac{3 G m^2}{R_* P} \ln \Lambda,
\end{equation}
to obtain
\begin{equation}
    \left| \frac{dP}{dt} \right| \gtrsim 9 \ln \Lambda \left( \frac{a}{R_*} \right) \left( \frac{m}{M_*} \right).
\end{equation}
This change in the period must be smaller than the observational limits $|\dot P|_{\rm max}$, so
\begin{align} \label{P_dot_constraint0}
    \frac{m}{M_\odot}  \lesssim & ~\frac{1}{9 \ln \Lambda} \left( \frac{R_*}{a} \right) 
    \left( \frac{M_*}{M_\odot} \right) 
    |\dot P|_{\rm max} \\
    \simeq &~ \frac{2 \times 10^{-18}}{\ln \Lambda} \left( \frac{R_*}{10 \, \mbox{km}} \right) \left( \frac{M_*}{M_\odot} \right)^{2/3} \left( \frac{P}{1300 \, \mbox{s}} \right)^{-2/3} \nonumber \\
    &~ \times \left( \frac{|\dot P|_{\rm max}}{3.6 \times 10^{-13}} \right), \nonumber
\end{align}
where we have used (\ref{a}) in the last estimate and have again scaled to quantities observed for GPMJ1839-10.  As we discussed above, we also know that the Hawking radiation limit gives
\begin{equation} \label{hawking}
    m \gtrsim 10^{-16} M_{\odot} 
\end{equation}
(see \cite{CarK20}), which we can combine with (\ref{P_dot_constraint0}) to obtain
\begin{align} \label{P_dot_constraint}
    \displaystyle
    R_* \gtrsim &~ 4.5 \times 10^8 \, \mbox{cm}  \left( \frac{M_*}{M_\odot} \right)^{-2/3} \left( \frac{\ln \Lambda}{10} \right) \left( \frac{P}{1300 \, \mbox{s}} \right)^{2/3} \nonumber \\ 
    &~ \left( \frac{|\dot P|_{\rm max}}{3.6 \times 10^{-13}} \right)^{-1}. 
\end{align}

\begin{figure}
    \centering
    \includegraphics[width = 0.45 \textwidth]{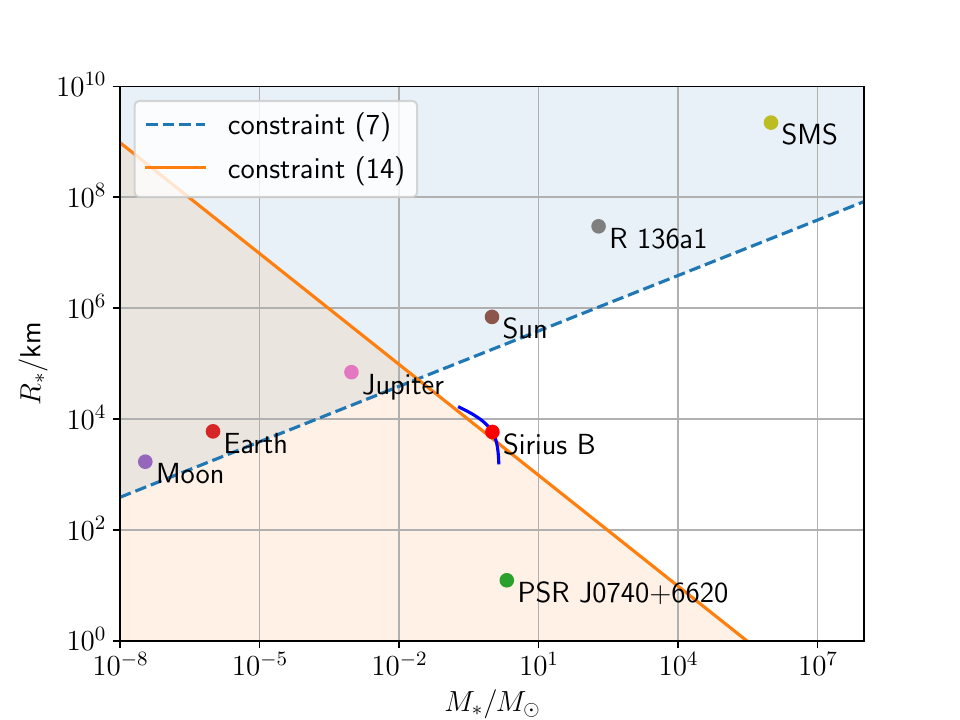}
    \caption{Constraints (\ref{P_constraint}) (dashed blue line) and (\ref{P_dot_constraint}) (solid orange line) are shown for the source of long-period radio transient GPMJ1839-10, assuming $\ln \Lambda = 10$.  In addition to typical solar-system objects we have included a massive neutron star (PSR J0740+6620, \cite{Miletal21}) and white dwarf (Sirius B, \cite{Bonetal17,HolBBCP98}), 
    one of the most massive observed stars (R 136a1, \cite{Kaletal22,Braetal22}), as well as a  hypothetical supermassive star (SMS) at the critical radius for collapse (see, e.g., \cite{ShaT83}).  The solid blue line shows the mass-radius relation for white dwarfs \cite{ZalMB10}.  Regions ruled out by the two constraints are shaded; only the unshaded triangle on the right part of the diagram satisfies both constraints.  We are not aware of any astrophysical objects that reside in the allowed region, although
    white dwarfs (e.g.~Sirius B) come close.}
    \label{fig:Fig1}
\end{figure}

If repeated passages of a PBH through a host star were to power LPRTs, then the star's radius $R_*$ would have to satisfy the two constraints (\ref{P_constraint}) and (\ref{P_dot_constraint}) simultaneously.  In Fig.~\ref{fig:Fig1} we draw both constraints, adopting the observed values for GPMJ1839-10 together with $\ln \Lambda = 10$, and shade those regions that are ruled out by the two constraints.   Only a triangle toward the right of the diagram satisfies both constraints.  As suggested by the figure, we are unaware of any astrophysical object that resides in this allowed region, even though white dwarfs come very close.

White dwarfs, however, are unlikely to be sufficiently effective in capturing PBHs at first place.  To see this, we first note that for Sirius B, for example, the constraint (\ref{P_dot_constraint0}) yields  $m \lesssim 10^{-16} M_\odot$, which, together with (\ref{hawking}), implies $m \simeq 10^{-16} M_\odot$.  Moreover, for a white dwarf we have $GM_*/(c^2 R_*) \simeq 10^{-4}$, so that, according to (\ref{energy_ratio}), energy loss is dominated by the drag term (\ref{deltaE}).  This energy loss results in capture if it exceeds the kinetic energy at infinite separation, $\Delta E_{\rm drag} > (1/2) \, m \varv_\infty^2$, or
\begin{equation}
    \left( \frac{\varv_\infty}{c} \right)^2 < 
    6 \ln \Lambda \left( \frac{G M_*}{c^2 R_*} \right) \left( \frac{M_*}{M_\odot} \right)^{-1} \left( \frac{m}{M_\odot} \right).
\end{equation}
Therefore, a white dwarf can capture a PBH of mass $m \simeq 10^{-16} M_\odot$ only if their relative speed at large separation satisfies $\varv_\infty \lesssim 10^{-9} c \simeq 3 \times 10^{-4} \, \mbox{km/s}$, which is significantly smaller than typical dispersion speeds in our Galaxy.  This means that a white dwarf is highly unlikely to capture a PBH with a mass as small as required by our constraints above (but see, e.g., \cite{PanL14,GenST20} for possible alternative capture mechanisms), making it unlikely that this process could act as the source of LPRTs.  

Even if the two constraints (\ref{P_constraint}) and (\ref{P_dot_constraint}) did identify a class of viable host stars that are sufficiently efficient in capturing PBHs in the required mass range, there would be further considerations to examine.  One of these concerns the liberated energy: assuming that the entire pulse energy originates from energy deposited by the PBH (rather than the PBH triggering the release of some other form of energy), the energy (\ref{deltaE}) would have to be at least as large as the observed values for the energy per pulse.  This provides a tighter constraint on $m$ that can be inserted in (\ref{P_dot_constraint0}) instead of (\ref{hawking}), resulting in an even more stringent constraint on $R_*$.  Moreover one would have to evaluate, of course, whether the specific characteristics of the observed radio emission could be explained in terms of PBHs traversing host stars. 

In summary, the discovery of LPRTs, whose long pulse periods challenges their interpretation as more traditional pulsars, encouraged us to examine whether these pulses might instead be caused by repeated passages of a PBH through a host star, and hence whether they might, in fact, provide clues on the nature of dark matter.  We demonstrate that observational data for the pulse period and its derivative provide constraints on the stellar radius that, when combined, rule out almost all known stellar objects as host stars.  As a remote possibility, a white dwarf seems least unlikely to act as a host star in this scenario, even though it is difficult to see how a white dwarf could capture as small a PBH as our estimates would require.  We therefore conclude that the capture of PBHs by a host star is highly unlikely to act as the source of LPRTs, but this does not rule out the existence of PBHs, of course, nor their capture by host stars.

\acknowledgements

This work was supported in parts by National Science Foundation (NSF) grants PHY-2010394 and PHY-2341984 to Bowdoin College, as well as NSF grants PHY-2006066 and PHY-2308242 to the University of Illinois at Urbana-Champaign.


%

\end{document}